\documentclass[12pt]{article}

\usepackage{graphics,epsfig}

\def\dalemb#1#2{{\vbox{\hrule height .#2pt
        \hbox{\vrule width.#2pt height#1pt \kern#1pt
                \vrule width.#2pt}
        \hrule height.#2pt}}}

\let\a=\alpha \let\b=\beta  \let\d=\delta \let\e=\epsilon
  \let\th=\theta  
\let\l=\lambda \let\m=\mu \let\n=\nu \let\x=\xi  
\let\s=\sigma   \let\f=\phi  
 
\let\w=\omega      \let\G=\Gamma  \let\Th=\Theta 
\let\X=\Xi  \let\S=\Sigma  \let\Y=\Psi
 
\let\la=\label  
  
\def\nn{\nonumber} \def\bd{\begin{document}} \def\ed{\end{document}}
\def\ds{\documentstyle} \let\fr=\frac \let\bl=\bigl \let\br=\bigr
\let\Br=\Bigr \let\Bl=\Bigl
\let\bm=\bibitem
\let\na=\nabla
\def\tU{{\widetilde U}}
\let\pa=\partial \let\ov=\overline
\def\ie{{\it i.e.\ }}
\newcommand{\be}{\begin{equation}}
\newcommand{\ee}{\end{equation}}
\def\ba{\begin{array}}
\def\ea{\end{array}}
\def\ft#1#2{{\textstyle{{\scriptstyle #1}\over {\scriptstyle #2}}}}
\def\fft#1#2{{#1 \over #2}}
\def\F#1#2{{ F_{#1}^{(#2)} }}
\def\cF#1#2{{ {\cal F}_{#1}^{(#2)} }}

\def\R{{\bf R}}
\def\sst#1{{\scriptscriptstyle #1}}
\def\oneone{\rlap 1\mkern4mu{\rm l}}
\def\e7{E_{7(+7)}}
\def\td{\tilde}
\def\wtd{\widetilde}
\def\im{{\rm i}}
\def\bog{Bogomol'nyi\ }
\newcommand{\ho}[1]{$\, ^{#1}$}
\newcommand{\hoch}[1]{$\, ^{#1}$}
\newcommand{\bea}{\begin{eqnarray}}
\newcommand{\eea}{\end{eqnarray}}
\newcommand{\ra}{\rightarrow}
\newcommand{\lra}{\longrightarrow}
\newcommand{\Lra}{\Leftrightarrow}
\newcommand{\ap}{\alpha^\prime}
\newcommand{\bp}{\tilde \beta^\prime}
\newcommand{\cB}{{\cal B}}
\newcommand{\cO}{{\cal O}}
\newcommand{\vecx}{\vec{x}}
\newcommand{\vecy}{\vec{y}}
\newcommand{\vecp}{\vec{p}}
\newcommand{\vecq}{\vec{q}}
\newcommand{\tr}{{\rm tr} }
\newcommand{\Tr}{{\rm Tr} }
\newcommand{\NP}{Nucl. Phys. }

\newcommand{\cL}{{\cal L}}
\newcommand{\cA}{{\cal A}}
\newcommand{\cD}{{\cal D}}

\def\sst#1{{\scriptscriptstyle #1}}
\def\0{{\sst{(0)}}}
\def\1{{\sst{(1)}}}
\def\2{{\sst{(2)}}}
\def\3{{\sst{(3)}}}
\def\4{{\sst{(4)}}}
\def\5{{\sst{(5)}}}
\def\6{{\sst{(6)}}}
\def\7{{\sst{(7)}}}
\def\8{{\sst{(8)}}}
\def\ve{\varepsilon}
\def\vf{\varphi}
\def\F{\Phi}
\def\wg{\wedge}

\newcommand{\tamphys}{\it 
}

\newcommand{\auth}{AUTHORS}

\def\thb{\bar{\theta}}
\def\Thb{\bar{\Theta}}
\def\barp{\bar{p}}
\def\barq{\bar{q}}
\def\barc{\bar{c}}
\def\bard{\bar{d}}
\def\e{\epsilon}

\def \bi{\bibitem}
\def \la {\label}

\def \l {\lambda}
\def\foot{\footnote}
\def \tl  {{\tilde \l}}
\def \sql {{\sqrt \l}}
\def \adss {$AdS_5 \times S^5$\ }
\newcommand{\rf}[1]{(\ref{#1})}
\def \ov {\over}

\def\th{\theta}
\def\Th{\Theta}
\def\vth{\vartheta}
\def\btheta{{\bar\theta}}
\def\ttheta{{{\tilde\theta}}}
\def\bttheta{{{\bar\ttheta}}}
\def\vth{\vartheta}

\def\ra{\rightarrow}
\def\N{{\cal N}}
\def\F{{\cal F}}
\def\uM{\underline{M}}
\def\uN{\underline{N}}
\def\uP{\underline{P}}
\def\cc{\circ}
\def\eqv{\equiv}

\def\ni{\noindent}

\def\Ep{E^{{}^{(+)}}}
\def\Em{E^{{}^{(-)}}}

\def\Mp{M^{{}^{(+)}}}
\def\Mm{M^{{}^{(-)}}}

\def \ha{{1\ov 2}}

\def\r{\rho}

\def\Y{{\rm Y}}
\def\X{{\rm X}}
\def\tY{\tilde{\rm Y}}
\def\tX{\tilde{\rm X}}
\def\dY{\dot{\rm Y}}
\def\dX{\dot{\rm X}}

\def \J {\mathcal{J}}
\def \del {\partial}

\def\dF{\dot{F}}
\def\dG{\dot{G}}
\def\df{\dot{f}}
\def \E {{\cal E}}
\def \S {{\cal S}}
\def \J {{\cal J}}

\def\ms{\mathcal{S}}
\def\mj{\mathcal{J}}
\def\soj{\fr{\ms}{\mj}}
\def \R {{\bf R}}
\def \om {\omega}
\def \bE {\bar E}
\def \x {{\cal X}}

\def \bi{\bibitem}
\def \la {\label}

\def \l {\lambda}
\def\foot{\footnote}
\def \tl  {{\tilde \l}}
\def \sql {{\sqrt \l}}
\def \adss {$AdS_5 \times S^5$\ }
\def \ov {\over}

\def \varpi {{\rm w}}

\def\thb{\bar{\theta}}
\def\Thb{\bar{\Theta}}
\def\psib{\bar{\psi}}
\def\barp{\bar{p}}
\def\barq{\bar{q}}
\def\barc{\bar{c}}
\def\bard{\bar{d}}
\def\e{\epsilon}

\def\At{\tilde{A}}
\def\Bt{\tilde{B}}
\def\ola{\overleftarrow}
\def\ora{\overrightarrow}
\def\at{\tilde{\a}}

\def\ps{\rlap{\, /}\;\,p }
\def\ks{\rlap{\, /}\;\,k }

\def\gym{g_{YM}}

\begin{document}
\overfullrule=0pt
\parskip=2pt
\parindent=12pt
\headheight=0in \headsep=0in \topmargin=0in
\oddsidemargin=0in

\vspace{ -3cm}
\thispagestyle{empty}

\begin{center}

{\Large\bf Toward getting finite results from $\N=4$ SYM
 \\
  with $\alpha'$-corrections   }

 \vspace{.5cm} { I.Y. Park\footnote{parkiy@longwood.edu } }\\
 \vskip 0.3cm

{\small\it Department of Chemistry and Physics, Longwood University\\
Farmville, VA 23909, USA }

\end{center}

 \vspace{0.1cm}

 \begin{abstract}
\ni  We take our first step toward getting finite results from the
$\alpha'$-corrected D=4 N=4 SYM theory with emphasis on the field
theory techniques. Starting with the classical action of the N=4 SYM
with the leading $\alpha'$-corrections, we examine new divergence at
one loop due to the presence of the $\alpha'$-terms. The new
vertices do not introduce additional divergence to the propagators
or to the three-point correlators. However they do introduce new
divergence, e.g., to the scalar four-point function which should be
canceled by extra counter-terms. We expect that the counter-terms
will appear in the 1PI effective action that is obtained by
considering the string annulus diagram. We work out the structure of
the divergence and comment on an application to the anomalous
dimension of the SYM operators in the context of AdS/CFT.

\end{abstract}
\newpage

\setcounter{equation}{0}
\setcounter{footnote}{0}
\setcounter{section}{0}


\section{Introduction}
In the recent developments of string theory the $D=4\;\; \N=4$ SYM
theory has played a much important role. The prominent example is
AdS/CFT correspondence where $\N=4$ SYM theory is employed to study
aspects of IIB supergravity/string theory on AdS$_5\times$S$_5$. The
$D=4\;\;\N=4$ theory approximates  an open superstring attached to a
set of D3-branes (see, e.g., \cite{Klebanov:1997bq} for a review) :
the results obtained by a full-fledged string computation will
reduce to those of the SYM in the $\a'\rightarrow 0$ limit. Since it
is a leading order approximation it may be worth studying a theory
that better approximates the open string than the pure SYM. We will
consider a classical action that is obtained by considering the open
string disc diagram, the $\a'$-corrected SYM. The action was
obtained in ten dimensions
\cite{Gates:1986is,Bergshoeff:1986jm,Tseytlin:1999dj,Cederwall:2001bt,
Bergshoeff:2001dc,Koerber:2001uu,Drummond:2004vf}. We keep the
leading $\a'$-correction terms which come at
$\a'^2$-order\footnote{In the literature
(e.g.,\cite{Koerber:2002zb,Stieberger:2006te,Medina:2006uf}) a few
higher orders were obtained as well for the bosonic sector. } and
reduce it to four
dimensions.\\

While $\N=4, D=4 $ SYM theory is a super-renormalizable theory the
status drastically changes once one adds the corrections from the
string theory since those correction terms are power-counting
non-renormalizable. The presence of the new vertices generates
additional divergence. In general even with a non-renormalizable
field theory one can consider an order-by-order renormalization, but
then the theory suffers from the loss of predictive power. This
would not be the case with the action of our starting point, the SYM
with $\a'$-corrections, since it comes from the string theory. As
well known open superstring yields finite results to various
scattering amplitudes, which are obtained via the world-sheet
technique. Therefore it may be worth seeing how the finiteness
results in the field theory context where divergence occurs. The
divergence would have to be cancelled by counter-terms.
Here we study the structures of the divergence and possible forms of
the counter-terms. It will be interesting to confirm (or disconfirm)
that the open string annulus diagram indeed implies the presence of
such terms. We leave the check to the future
string theory based computation \cite{progress}.  \\

The remainder of the paper is organized as follows. In sec2, we
consider the SYM plus the $\a'^2$-results that are obtained  in the
literature \cite{Cederwall:2001bt,
Bergshoeff:2001dc,Koerber:2001uu}. These are ten dimensional
results: we carry out dimensional reduction to four dimensions. Out
of the terms that result we record only the terms that are relevant
for our computation. More complete expression is presented in
Appendix B. With the dimensional regularization we examine, at
one-loop and $\a'^2$ order, various divergence. We note that the new
vertices do not introduce any new divergence to the propagators due
to an identity concerning the scaleless integrals in the dimensional
regularization. We move to new three point correction graphs, which
vanish as well. With the three point correlators it is the color
index structure that makes them vanish. Non-vanishing divergence
appears with four-point functions. We take the example of the scalar
four-point functions and work out the divergent parts of the
integrals. We then look into the possible forms of the
counter-terms. We illustrate this with an example. Section 3 has
discussions of issues that are related to the current computation.
We also comment on future directions. In Appendix A, we present our
notations and conventions for the SYM, and list the Z-factors of the
wave-function renormalization.

 \section{New divergence from sringy vertices}

The {\N=4} action with the $\a'$-corrections is quoted in the
appendices along with our conventions. Below we consider, at
one-loop and $\a'^2$-order, new graphs to two-, three- and the
 four- point functions that are introduced by the stringy
vertices. In the case of the four point function we only consider
the scalar external lines. The new graphs of two and three point
function vanish, but as for the four point graphs one has
non-vanishing results. We analyze their structure and discuss the
counter terms that remove the divergence.

\subsection{propagators and three-point functions}

The stringy vertices produce new graphs of radiative corrections to
the propagtors. We present a few of them in Fig.1 below. They (and
all the other propagator corrections at $\a'^2$-order) vanish due to
an identity concerning the scaleless integrals in the dimensional
regularization (see, e.g.,\cite{collins,sterman}),
 \bea
 \int d^dq\;( q^2)^\b=0 \label{qid}
 \eea
where $\b$ is an arbitrary number.

\begin{figure}[!ht]
\centerline{
        \begin{minipage}[b]{12cm}
               \epsfxsize=10
               cm
                \epsfbox{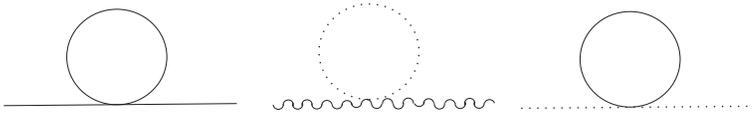}
        \end{minipage}
} \caption{Examples of the new graphs for the propagators   }
\label{2pt}
\end{figure}

 The one loop corrections to the three point
correlators also vanish: some of them for the same reason as the
propagator corrections. There are other graphs that do not contain
the scaleless integrals. They have two vertices, one from the pure
SYM and the other from the $\a'^2$-vertices as illustrated in Fig.2.
These graphs vanish because of their color structure: they all come
with
 \bea
 \sim f^{def}\; \mbox{Str}(T^eT^f\cdots) =0
 \eea
where $T$'s are the SU(N) group generators in the adjoint
representation, ${T^b}_{ac}=if^{abc}$
\begin{figure}[!ht]
\centerline{
        \begin{minipage}[b]{12cm}
               \epsfxsize=10cm
                \epsfbox{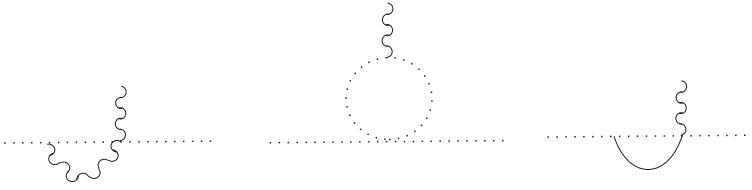}
        \end{minipage}
} \caption{Examples of the new graphs for the three point function
} \label{3pt}
\end{figure}

\subsection{ four-scalar vertices}

At one-loop, there are altogether five graphs shown in Fig.3. In
each graph one vertex comes from the $\a'^2$-terms and the other(s)
from the pure SYM part. The stringy vertices are presented in
Appendix B. All the graphs contain the common factor of
 \bea
 (2\pi\a')^2\gym^8f^{mea}f^{mfb}\;\mbox{Str}({T^eT^fT^cT^d})\;
 \fr{(2\pi)^4\d({\sum_{k=1}^4}p_k)}{p_1^2p_2^2p_3^2p_4^2}\fr{\G(2-\w)}{(4\pi)^2}
 \label{cf}
 \eea
where $\w\equiv D/2$. Our conventions are explained in Appendix A.
\begin{figure}[!ht]
\centerline{
        \begin{minipage}[b]{9cm}
               \epsfxsize=9cm
                \epsfbox{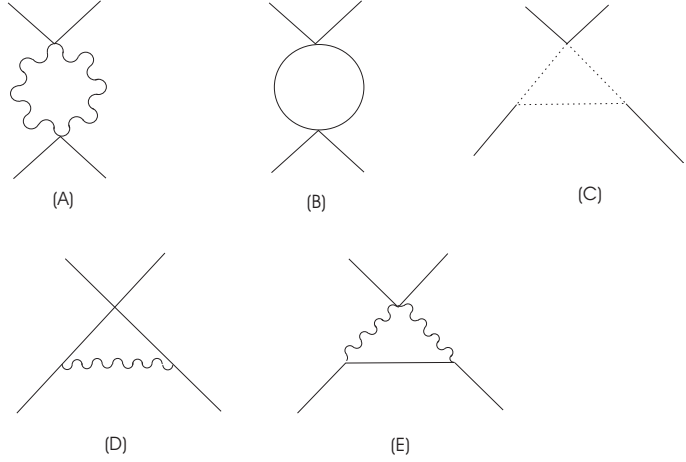}
        \end{minipage}
} \caption{New graphs for the scalar four point function   }
\label{3pt}
\end{figure}
We summarize our results as follows. One of the two vertices in
graph (A), with the other being one of the standard SYM vertices,
comes from
 \bea
(2\pi\a')^2\;\;\mbox{Str}\left[
-\fr{1}{8}F_{\m\n}F^{\m\n}D_\r\phi_kD^\r\phi^k
 -\fr{1}{2} D_\n\phi_iF_{\n\r}F^{\r\s}D^\s\phi^i \right]
 \eea
Here and below only the regular partial derivative part of the
covariant derivative will contribute. The sum of these two
contributions is given by\footnote{ "$\Rightarrow$" indicates the
fact that only the divergent parts have been recorded. The results
are given in the momentum space obtained by Fourier transformation.
Here and below the perm stands for the terms that are obtained by
permutations of
 \[
\{ (p_1,a,i),(p_2,b,j),(p_3,c,k),(p_4,d,l)\}
 \]
 }
 \bea
\!\!\!\!&&
<\phi_i^a(x_1)\phi_j^b(x_2)\phi_k^c(x_3)\phi_l^d(x_4)>_{_{(A)}}
 \Rightarrow \d_{ij}\d_{kl}\left[\fr{1}{24}(p_1+ p_2)^2(p_3\cdot p_4)
 -\fr1{6}\;\;(p_1+p_2)\!\cdot\! p_3\;(p_1+p_2)\!\cdot\! p_4
 \right]\nn\\
 &&\hspace{5.5in}+\mbox{perm} \label{a}
 \eea
The stringy vertex in graph (B) come from
 \bea
(2\pi\a')^2\;\;\mbox{Str}\left[-\fr{1}{8} D_\m\phi_j D^\m\phi^j
D_\n\phi_kD^\n\phi^k+\fr{1}{4}D_\n\phi_iD^\n\phi^kD_\s\phi_kD^\s\phi^i
 \right]
 \eea
which yields
 \bea
 <\phi_i^a(x_1)\phi_j^b(x_2)\phi_k^c(x_3)\phi_l^d(x_4)>_{_{(B)}}
\Rightarrow && \d_{ij}\d_{kl}\left[\fr16(p_1+p_2)^2\;p_3\cdot p_4
 -\fr5{12}\;(p_1+p_2)\!\cdot\! p_3\;(p_1+p_2)\!\cdot\! p_4
 \right]\nn\\
&&+\d_{ik}\d_{jl}\left[\fr1{4}\;(p_1+p_2)^2\;p_3\cdot p_4
\right]+\mbox{perm}\label{b}
 \eea
For graph (C) the relevant terms are
 \bea
(2\pi\a')^2\;\;\mbox{Str}\left[\fr{1}{4}\psib\G_\m D_\n\psi
D^\m\phi^iD^\n\phi_i -\fr{1}{4}\psib\G_{\m nk}D_\s\psi
D^\m\phi^nD^\s\phi \right]
 \eea
 The result for graph (c) as it
comes out of the computation is
 \bea
 <\phi_i^a(x_1)\phi_j^b(x_2)\phi_k^c(x_3)\phi_l^d(x_4)>_{_{(C)}}
 \Rightarrow &&\d_{ij}\d_{kl}[-\fr13(p_1+p_2)^2\;p_3\cdot p_4
 -\fr2{3}\;(p_1+p_2)\!\cdot\! p_3\;(p_1+p_2)\!\cdot\! p_4\nn\\
&&\quad\quad-(p_1\cdot p_2)(p_3\cdot p_4)+(p_1\cdot p_3)(p_2\cdot
p_4) -(p_1\cdot p_4)(p_2\cdot p_3)\nn\\
&&\hspace{1in}\quad\quad-2(p_2\cdot p_3)(p_2\cdot p_4)]
\nn\\
 &&+\d_{ik}\d_{jl}\left[2(p_1\cdot p_3)(p_2\cdot p_4)-
 2(p_1\cdot p_4)(p_2\cdot p_3)\right]+\mbox{perm}\nn
 \eea
It can be simplified by utilizing the SO(6)- and color- index
structures:
 \bea
 <\phi_i^a(x_1)\phi_j^b(x_2)\phi_k^c(x_3)\phi_l^d(x_4)>_{_{(C)}}
 \Rightarrow &&\d_{ij}\d_{kl}[-\fr13(p_1+p_2)^2\;p_3\cdot p_4
 -\fr2{3}\;(p_1+p_2)\!\cdot\! p_3\;(p_1+p_2)\!\cdot\! p_4\nn\\
&&-(p_1\cdot p_2)(p_3\cdot p_4)-(p_1\cdot p_3)(p_1\cdot p_4)
-(p_2\cdot p_3)(p_2\cdot p_4)
 ]\nn\\
 &&+\d_{ik}\d_{jl}\left[2(p_1\cdot p_3)(p_2\cdot p_4)-
 2(p_1\cdot p_4)(p_2\cdot p_3)\right]+\mbox{perm}\nn\\ \label{c}
 \eea
The stringy vertex for graph (D) is the same as that of (B). One
gets the following result,
 \bea
<\phi_i^a(x_1)\phi_j^b(x_2)\phi_k^c(x_3)\phi_l^d(x_4)>_{_{(D)}}
 \Rightarrow &&\d_{ij}\d_{kl}\left[-\fr{5}{8}(p_1\cdot p_2)(p_3\cdot p_4)
 -\fr1{12}\;p_1^2(p_3\cdot p_4)-\fr18p_2^2(p_3\cdot
 p_4)\right.\nn\\
&&\left.\quad\quad\quad+\fr1{12}(p_1\cdot p_3)(p_1\cdot
p_4)+\fr14(p_1\cdot p_4)(p_2\cdot
p_3) \right] \nn\\
&& +\d_{ik}\d_{jl}\left[-\fr14(p_1\cdot p_3)(p_2\cdot p_4)+
 \fr14(p_1\cdot p_4)(p_2\cdot p_3)\right.\nn\\
 &&\left.
 \quad\quad\quad+\fr12p_2^2(p_3\cdot p_4)+\fr14p_1^2(p_3\cdot p_4)
 +\fr94(p_1\cdot p_2)(p_3\cdot p_4)\right]\nn\\
 &&+\mbox{perm}\nn
 \eea
which can be rewritten, by utilizing the SO(6)- and color- index
structures, as
 \bea
<\phi_i^a(x_1)\phi_j^b(x_2)\phi_k^c(x_3)\phi_l^d(x_4)>_{_{(D)}}
 \Rightarrow &&\d_{ij}\d_{kl}\left[-\fr{5}{12}(p_1\cdot p_2)(p_3\cdot p_4)
 -\fr{5}{48}\;(p_1+p_2)^2(p_3\cdot p_4)\right.\nn\\
&&\left.\quad\quad\quad+\fr1{12}(p_1\cdot p_3)(p_1\cdot
p_4)+\fr18(p_1\cdot p_3)(p_2\cdot p_4)\right.\nn\\
&& \left.\quad\quad\quad+\fr18(p_1\cdot p_4)(p_2\cdot
p_3) \right] \nn\\
&& +\d_{ik}\d_{jl}\left[-\fr14(p_1\cdot p_3)(p_2\cdot p_4)+
 \fr14(p_1\cdot p_4)(p_2\cdot p_3)\right.\nn\\
 &&\left.
 \quad\quad\quad+\fr12p_2^2(p_3\cdot p_4)+\fr14p_1^2(p_3\cdot p_4)
 \right.\nn\\
 &&\left.\quad\quad\quad+\fr94(p_1\cdot p_2)(p_3\cdot p_4)\right]+\mbox{perm}
 \label{d}
 \eea
Finally the graph (E), whose stringy vertex is the same as that of
(A), yields vanishing result:
 \bea
 <\phi_i^a(x_1)\phi_j^b(x_2)\phi_k^c(x_3)\phi_l^d(x_4)>_{_{(E)}}
\Rightarrow &&0 \label{e}
 \eea
Summing up (\ref{a})-(\ref{d}) one gets
 \bea
<\phi_i^a(x_1)\phi_j^b(x_2)\phi_k^c(x_3)\phi_l^d(x_4)>_{_{\mbox{total}}}
 \Rightarrow&&\d_{ij}\d_{kl}\left[-\fr{11}{48}\;(p_1+p_2)^2(p_3\cdot p_4)
 -\fr54\;\;(p_2\!\cdot\! p_3)(\;p_2\!\cdot\! p_4)\right.\nn\\
 &&\left.\quad\quad-\fr{5}{12}(p_1\cdot p_2)(p_3\cdot p_4)
 -\fr{7}{6}(p_1\cdot p_3)(p_1\cdot
p_4)\right.\nn\\
 && \left.\quad\quad\quad-\fr98(p_1\cdot p_3)(p_2\cdotp_4)
 -\fr98(p_1\cdot p_4)(p_2\cdot p_3) \right] \nn\\
&& +\d_{ik}\d_{jl}\left[+\fr74(p_1\cdot p_3)(p_2\cdot p_4)-
 \fr74(p_1\cdot p_4)(p_2\cdot p_3)\right.\nn\\
 &&\left.
 \quad\quad\quad+\fr34p_2^2(p_3\cdot p_4)+\fr12p_1^2(p_3\cdot
 p_4)\right.\nn\\
 &&\left.\quad\quad\quad
 +\fr{11}4(p_1\cdot p_2)(p_3\cdot p_4)\right]+\mbox{perm}
 \label{total}
 \eea
The counter-terms that remove the divergence can readily be
obtained. We illustrate this with
$\d_{ik}\d_{jl}\;\fr{11}{4}(p_1\cdot p_2)(p_3\cdot p_4)$-term in
(\ref{total}). Including the common factor (\ref{cf}) it is
 \bea
 (2\pi\a')^2\gym^8f^{mea}f^{mfb}\;\mbox{Str}({T^eT^fT^cT^d})\;
 \fr{(2\pi)^4\d({\sum_{k=1}^4}p_k)}{p_1^2p_2^2p_3^2p_4^2}\fr{\G(2-\w)}{(4\pi)^2}
  \d_{ik}\d_{jl}\;\fr{11}{4}(p_1\cdot p_3)(p_2\cdot p_4)\nn\\
 \eea
It can be removed by adding the following counter-term in the action
 \bea
-\fr{11}{4}(2\pi\a')^2\gym^2
f^{mep}f^{mfq}\;\mbox{Str}({T^eT^fT^cT^d})\;
 \fr{\G(2-\w)}{(4\pi)^2}\;\pa_\m\phi_m^p \pa_\m\phi_n^q
\pa_\n\phi_m^g \pa_\n\phi_n^h
 \eea
The counter-terms for other parts of the divergence can be similarly
determined.

\section{Discussions and Future Directions}

One of the reasons why the present computation may be useful is the
fact that a D-brane is a stringy object: it will take the full open
string theory for a complete description of the object. The methods
of the description of a D-brane are at the heart of the AdS/CFT. The
relevance of the open string in the context of AdS/CFT was discussed
e.g., in \cite{Park:1999xz, Park:2001bm,Park:2000du}.\footnote{
Related discussions may be found in
 \cite{DiVecchia:2005vm,Bertolini:2001qa}.} The leading
approximation of the open superstring is the SYM theory. Although
simple and useful it does not contain the effects of the massive
open string modes. Therefore it may be meaningful to try to
accommodate them. There are two ways to do that. First one may turn
to the world-sheet description for various scattering amplitudes. At
a given loop order, it will include the complete effects of the
massive modes. Less inclusive but still advantageous in other
aspects is the regular field theory approach. Efficient to include
the massive modes, the world-sheet theory does not have the same
status as a regular field theory since string field theory is less
developed although there has been some progress
\cite{Rastelli:2001uv,Bars:2002qt}. Furthermore, unlike the abelian
case where the effective action can be obtained in a closed form
(see, e.g., \cite{Tseytlin:1998kw} for a relatively recent
discussion), in the non-abelian case one must consider four-point,
five-point, etc, separately, and deduce the field theory action from
the results. It may be useful for that purpose to know the possible
forms of the field theory counter-terms in advance through an
analysis such as the present one. In other words, the string-based
technique and the field theory technique may be mutually
guiding.\footnote{Related discussions for the pure SYM case can be
found in \cite{Refolli:2001df,Grasso:2002wb}}.\\

We comment on two potential applications of our results. In the
literature, there have been pieces of evidence
\cite{Chepelev:1997fk,Keski-Vakkuri:1997pr,Gonzalez-Rey:1998uh} that
the perturbative quantum corrections of pure SYM theory can be
mapped to the terms in the DBI action in the AdS$_5\times$S$^5$
curved background. ( Related discussions can be found in
\cite{Kuzenko:2002sv,Kuzenko:2002xq}.) Once we complete the check of
the counter-terms through the string analysis we will be in a
position to see how they would modify the story. Presumably they
would not change the big picture but only some details such as the
field redefinition introduced in \cite{Gonzalez-Rey:1998uh}. The
other application is that one may investigate whether/how the
$\a'$-terms correct the anomalous dimensions of the SYM operators
that appear in the context of AdS/CFT \cite{Minahan:2002ve}.

We end with a few side remarks. One way to interpret the results of
\cite{Chepelev:1997fk,Keski-Vakkuri:1997pr,Gonzalez-Rey:1998uh} is
that putting the action in the curved background amounts to having
the 1PI effective action, $\G$. In other words, although one starts
out with the SYM (or open string) in a flat space the theory
completes itself in the curved target space(in the sense of a
non-linear sigma type model). The advantage of having the 1PI action
handy is that one only computes the tree graphs since the action
already contains all the quantum corrections. Therefore to compute
certain physical quantities one can either start with the flat space
action and include the quantum corrections, or alternatively use the
1PI action, which would be equivalent to using the action in the
curve space\footnote{\ni The effective action in
\cite{Gonzalez-Rey:1998uh} was obtained by S-dualizing the SYM
one-loop effive action, hence would contain not only the
perturbative quantum effect but also the non-perturbative effect
although there is some subtlety as expressed in
\cite{Gonzalez-Rey:1998uh}. Therefore it is more than a 1PI action
since typically a 1PI action refers only to the parts that are
obtained by perturbative techniques.}, and compute the tree graphs.
However, for certain purposes such as mechanically finding the SYM
operators that are dual to the supergravity modes
\cite{Das:1998ei,Park:2000du,Nurmagambetov:2001ab} or implementing
the duality at a lagrangian-to-lagrangian level
\cite{Park:2001bm}\footnote{The computational techniques are curved
space generalization of those of \cite{Gibbons:2000hf,Sen:2000kd}. }
it seems to take the action in the curved background from the
beginning.

\vspace{2in}

\section*{Acknowledgments }

I am grateful to M. Rocek and G. Sterman for their valuable
discussions in the various stages of this work. Part of the results
of the work was presented in  KIAS string theory workshop, summer
2006. I thank for their hospitality during my stay.

\newpage

\renewcommand{\theequation}{A.\arabic{equation}}
 \setcounter{equation}{0}
  \section*{Appendix A: Notations and Conventions
   }
$\N=4$ SYM action with the leading string correction is given by
 \bea
  {\cal L}= {\cal L}_{SYM}+ {\cal L}_c
 \eea
with
 \bea
  {\cal L}_{SYM}&=&\left[
   -\fr14 F^a_{\m\n}F^{a\m\n}
 -\fr12 \left(\pa_\m\phi_i^a+f^{abc}A_\m^b\phi_i^c\right)^2
  -\fr12\bar{\psi}^a\G^{\m}\left(\pa_\m\psi^a+f^{abc}A_\m^b\psi^c\right)
  \right.\\
  && \left. -\fr12\;f^{abc}\;\psib^a\G^i\phi_i^b\;\psi^c
  -\fr14 \sum_{i,j}f^{abc}f^{ade}\phi_i^b\phi_j^c\phi_i^d\phi_j^e
   -\fr12\pa_\m\w_a^*\left(\pa^\m\w_a+f^{abc}A_b^\m\w_c\right)\right]
 \nn
 \eea
 where $\psi$ is a thirty two component Mayorana-Weyl spinor with four dimensional
 space-time dependence. The conjugation is is defined by
 \bea
 \psib\equiv  \psi^\dagger i\G^0
 \eea
 The $\a'^2$-order  terms in ${\cal L}_c$
 (which is the leading correction) are given in Appendix B.
To take into account the fact that $\psi$ is a Mayorana-Weyl spinor
one uses the following relation \cite{Erickson:2000af} at the end of
the trace algebra,
 \bea
\tr\; \G^\m\G^\n=16 \d^{\m\n}
 \eea
 The Z-factors of the wave-function
renormalization are as follows:
 \bea
 Z_\phi=1+\fr{\l}{8\pi^2}\G(2-w)\quad\quad
 Z_\psi=1+\fr{4\l}{16\pi^2}\G(2-w)\quad\quad
Z_A=1+\fr{\l}{8\pi^2}\G(2-w)\nn\\
 \label{zphiMZ'}
 \eea
The first two Z-factors are given, e.g., in \cite{Erickson:2000af}.

\renewcommand{\theequation}{B.\arabic{equation}}
 \setcounter{equation}{0}
  \section*{Appendix B: Dimensional reduction of the leading
  $\a'$-corrections
   }
In $D=10$ Minkowski space the $\N=1$ SYM action with leading string
corrections \cite{Gates:1986is,Bergshoeff:1986jm,Cederwall:2001bt,
Bergshoeff:2001dc,Koerber:2001uu} is
 \bea
 {\cal L}_{\a'^2,D=10}&=&\mbox{Str}\;\;
 (2\pi)^2\a'^2\left[\fr{1}{8}F^{MN}F_{NP}F^{PQ}F_{QM}
 -\fr{1}{32}\left(F^{MN}F_{MN}\right)^2\right.\nn\\
 &&\left.-\fr{1}{4}\;\psib\G_MD_N\psi\;F^{MP}{F_P}^N
 +\fr{1}{8}\;\psib\G_{MNP}D_Q\psi\,F^{MN}F^{PQ}
 +\fr{1}{24}\;\psib\G^{M}D^{N}\psi\,\psib\G_MD_N\psi\right.\nn\\
&&\left.+\,\fr{7}{480}F_{MN}\;\psib\G^{MNP}\psi\,\{\psib,\G_P\psi\}
-\,\fr{\at^2}{2880}F_{MN}\;
\psib\G_{PQR}\psi\,\{\psib,\G^{MNPQR}\psi\}\right]
 \eea
where Str denotes the symmetrized trace on color indices which are
suppressed,
 \bea
 \mbox{Str}\;A_1A_2\cdots A_n=\fr{1}{n!}\tr\left(  A_1A_2\cdots
 A_n+\mbox{all permutations}\right)
 \eea
Keeping the terms with up to two fermion fields, the dimensionally
reduced action is as follows:

 \bea
 \fr{{\cal L}_{\a'^2,D=4}}{(2\pi)^2\a'^2}=&&
 -\fr{1}{32}\left(F_{\m\n}F^{\m\n}F_{\r\s}F^{\r\s}+
 [\phi_i,\phi_j][\phi^i,\phi^j][\phi_k,\phi_l][\phi^k,\phi^l]
+4D_\m\phi_j D^\m\phi^j D_\n\phi_kD^\n\phi^k \right.\nn\\
 && \left.\quad\quad  -2F_{\m\n}F^{\m\n}[\phi_i,\phi_j][\phi^i,\phi^j]
 +4F_{\m\n}F^{\m\n}D_\r\phi_kD^\r\phi^k-4[\phi_i,\phi_j][\phi^i,\phi^j]
D_\m\phi_kD^\m\phi^k  \right)\nn\\
&& +\fr{1}{8}\left(
 F_{\m\n}F^{\n\r}F_{\r\s}F^{\s\m}
 +[\phi_i,\phi_j][\phi^j,\phi^k][\phi_k,\phi_l][\phi^l,\phi^i]
-4D_\n\phi_iF_{\n\r}F^{\r\s}D^\s\phi^i\right.\nn\\
 && \left.\quad\quad+2D_\n\phi_iD^\n\phi^kD_\s\phi_kD^\s\phi^i
 +4iD_\n\phi_iF^{\n\r}D_\r\phi_l[\phi^l,\phi^i]
 +4D_\n\phi_iD^\n\phi^k[\phi_k,\phi_l][\phi^l,\phi^i]\right)\nn\\
&&-\fr14\left(\psib\G_\m D_\n\psi F^{\m\r}{F_\r}^\n
 -\psib\G_\m D_\n\psi D^\m\phi^iD^\n\phi_i
 -i\psib\G_\m [\phi_i, \psi]F^{\m\n}D_\n\phi^i \right.\nn\\
 && \left.\quad\quad+\psib\G_i D_\m\psi F^{\m\n}D_\n\phi^i
 +\psib\G_\m [\phi_i, \psi]D^{\m}\phi^j[\phi^i,\phi_j]
 -i\psib\G_iD_\m\psi D^{\m}\phi^j[\phi^i,\phi_j]\right.\nn\\
 && \left.\quad\quad+i\psib\G_i [\phi_j,\psi] D^\m\phi^iD_\m\phi_j
 +i\psib\G_i [\phi_j,\psi][\phi^i,\phi^k][\phi_k,\phi^j]
 \right)\nn\\
 &&+\fr18\left(\psib\G_{\m\n\r}D_\s\psi F^{\m\n}F^{\r\s}
-i\psib\G_{\m\n\r}[\phi_l,\psi]F^{\m\n}D^{\r}\phi^l
 -\psib\G_{\m\n k}D_\s\psi F^{\m\n}D^\s\phi^k \right.\nn\\
 &&\left.\quad\quad -2\psib\G_{\m\r n}D_\s\psi D^\m\phi^n F^{\r\s}
 -\psib\G_{\m\n k}[\phi_l,\psi]F^{\m\n}[\f^k,\f^l]
 +2i\psib\G_{\m\r n}[\phi_l,\psi]D^\m\phi^nD^\r\phi^l \right.\nn\\
 &&\left.\quad\quad-2\psib\G_{\m  nk}D_\s\psi D^\m\phi^nD^\s\phi^k
 -2\psib\G_{\m nk}[\phi_l,\psi]D^\m\phi^n[\f^k,\f^l]\right.\nn\\
 &&\left.\quad\quad-\psib\G_{mn\r}[\phi_l,\psi][\f^m,\f^n]D^\r\phi^l
 -i\psib\G_{mn\r}D_\s\psi[\f^m,\f^n]F^{\r\s}\right.\nn\\
 &&\left.\quad\quad+i\psib\G_{mnk}D_\s\psi[\f^m,\f^n]
 D^\s\phi^k
 +i\psib\G_{mnk}[\phi_l,\psi][\f^m,\f^n][\f^k,\f^l]
 \right)+\cdots
 \eea
where $\psi$ is a thirty two component Mayorana-Weyl spinor with
four dimensional space-time dependence.

\newpage

\end{document}